\documentclass[%
 reprint,
 amsmath,amssymb,
 aps,
]{revtex4-1}

\usepackage{graphicx}
\usepackage{dcolumn}
\usepackage{bm}

\begin{document}

\preprint{APS/123-QED}

\title{Observation of Mollow quintuplet in $F=3/2$ hyperfine structure state of $^{3}\textrm{He}$ atomic cell}

\author{Yuanzhi Zhan}
\author{Xiang Peng}%
\author{Sheng Li}
\author{Liang Zhang}%
\author{Jingbiao Chen}%
\author{Hong Guo}
 \email{hongguo@pku.edu.cn}

\affiliation{%
 State Key Laboratory of Advanced Optical Communication Systems and Networks, School of Electronics Engineering and Computer Science, and Center for Quantum Information Technology, Peking University, Beijing 100871, China
}%

\date{\today}

\begin{abstract}
We experimentally observed the Mollow quintuplet (MQ) in $F=3/2$ hyperfine structure state of $^{3}\textrm{He}$ atoms. The metastability-exchange collisions (MECs) transfer the Mollow Triplet (MT) from the ground states of $^{3}\textrm{He}$ atoms to the metastable states, and the MQ is demonstrated by four Zeeman levels of $F=3/2$ hyperfine states with linearly polarized light. The similar effect also achieves in the mixture cell of $^{3}\textrm{He}$ and $^{4}\textrm{He}$.
\end{abstract}

\pacs{Valid PACS appear here}
\maketitle

The interaction between light and materials is an important research area of quantum system. The Fermi Golden Rule describes the weak coupling between the light and atoms, which reveals the transition between the energy eigenstates of atoms \cite{fermi1950}. Mollow triplet (MT) is described as that two-level atoms are driven by strong coherent field in free space \cite{mollow1969power}, which reveals the change from singlet spectrum to the triplet. MT of resonant light scattering was achieved in the atomic beam of sodium \cite{wu1975investigation}, quantum dot \cite{fischer2016self}, silicon vacancy of diamond \cite{zhou2017coherent}, and superconducting circuits \cite{baur2009measurement}. The ratio of center peak and sidebands of MT is influenced by coherence \cite{nathan2015theory} or collisions \cite{khoa2016mollow}. The quintuplet spectrum has been reported in three different systems. The first one is three-level atoms coupled with two high-intensity resonant laser beam sharing a common level \cite{cohen1977simultaneous}, the second one is two-level atoms coupled with bimodality cavity \cite{Nha2000resonance}, and the third one is the resonance fluorescence spectrum of quantum dots system \cite{vam2009spin}. The so-called Mollow quintuplet (MQ) is modeled by summing Mollow triplets (MTs) of two natural excitons polarized in orthogonal direction of quantum dot with the same laser \cite{rong2013mollow}, and furthermore two modified MTs can induce the Mollow nonuplets (MNs) in two coupled quantum dots system at strong exchange regime \cite{angelatos2015entanglement}. 

\begin{figure}
	\includegraphics[width=0.4\textwidth]{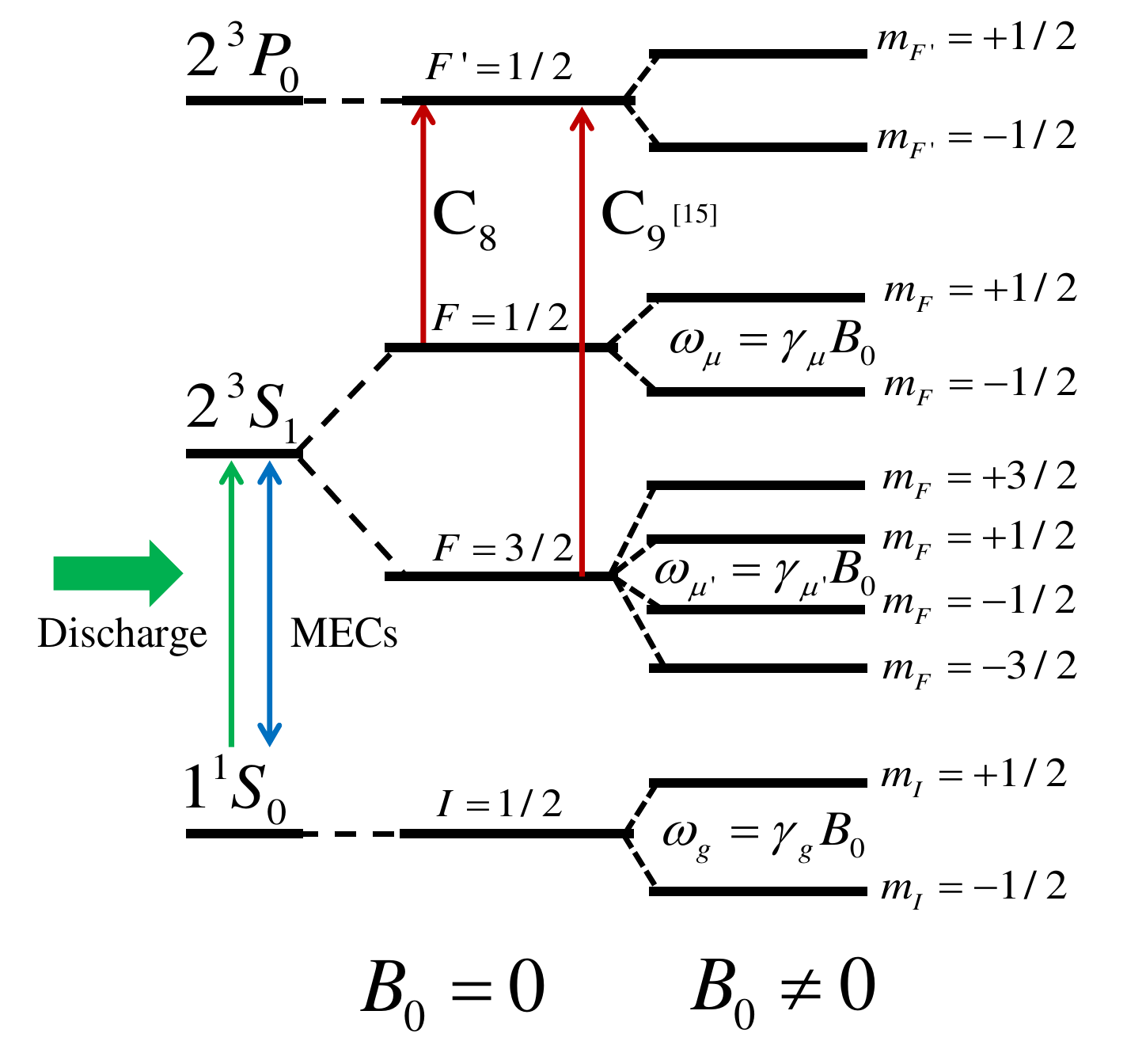}
	\caption{\label{fig:energy} The energy-level diagram of $^{3}\textrm{He}$ atoms (not in scale). $1^{1}S_{0}$, $2^{3}S_{1}$, $2^{3}P_{0}$ are the ground state, metastable state, excited state, respectively. The metastable state contains two hyperfine states, $F=1/2$ and $F=3/2$, each of which is split into 2 or 4 Zeeman sublevels in a static magnetic field $\bf{B_0}$. The blue line indicates the metastability exchange collisions (MECs) between $1^{1}S_{0}$ and $2^{3}S_{1}$. The interval depends on the gyromagnetic ratio, i.e. $\gamma_{\rm g}=$~3.2~kHz/G for ground state, $\gamma_\mu=$~3.8~MHz/G and $\gamma_{\mu'}=$~1.9~MHz/G for $F=1/2$, $F=3/2$, respectively \cite{colegrove1963polarization}. The green line indicates the RF discharge to generate $2^{3}S_{1}$ atoms. The red line indicates the optical transition $\textrm{C}_8$ between the $2^{3}S_{1}$ and $2^{3}P_{0}$ states with the vacuum wavelength 1083.353 nm, which is used for optical pumping and probing \cite{nacher1985optical}.
	}
\end{figure}

In this letter we observe the MQ in $F=3/2$ hyperfine structure state of $^{3}\textrm{He}$ atoms by detection with linearly polarized light, and the experimental results indicate to the alignment effect with a different physical process from that mentioned before. We use a RF field to drive the ground state $1 ^1S_0$, and the MT is induced in $m_I=+1/2$ and $m_I=-1/2$ states, according to the energy-level diagram of $^{3}\textrm{He}$ atoms shown in Fig.~\ref{fig:energy}. The metastability-exchange collisions (MECs) can transfer the MT from the ground states to $F=1/2$ metastable states \cite{zhan2018observation}, and the transferred MT also induced the further MQ in $F=3/2$ metastable states. The experimental results reveal that the MQ effect is related the higher order spin quantum number state and the detection with linearly polarized light.

\begin{figure}
	\includegraphics[width=0.45\textwidth]{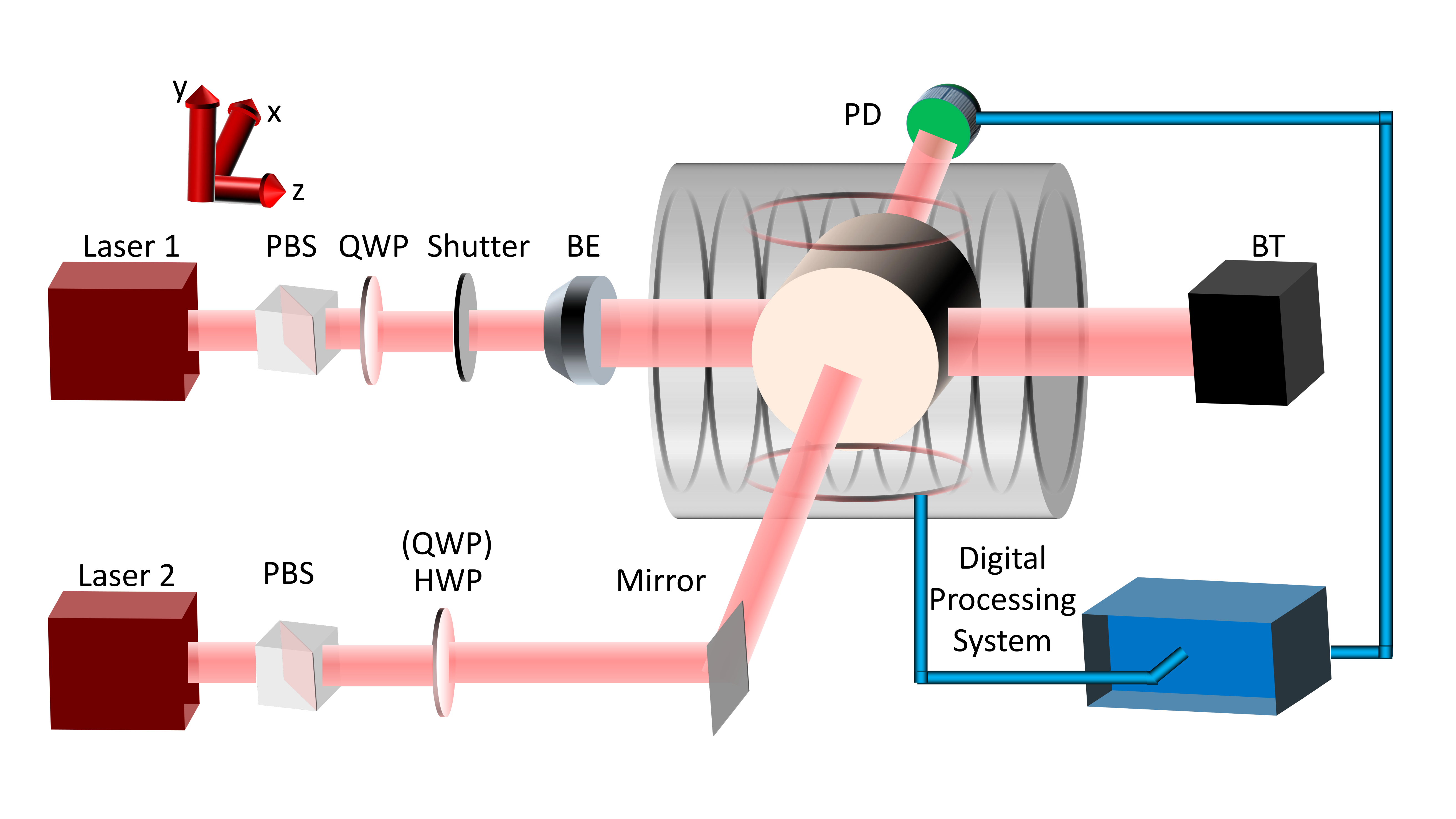}
	\caption{\label{fig:experiment} The experimental setup for the MQ measurement in $^{3}\textrm{He}$ atoms. PBS: Polarization Beam Splitter, QWP: Quarter Wave Plate, HWP: Half Wave Plate, BE: Beam Expander, BT: Beam Trap, PD: Photo Detector.
	}
\end{figure}

The experimental setup is shown in Fig.~\ref{fig:experiment} and we use the two lasers to pump and detect $F=3/2$ hyperfine states of metastable states, respectively. Both the pump (laser 1) and the probe (laser 2) beams are from laser source (NKT Photonics Y10), and the pump beam is power-enhanced by the a laser amplifier (LEA Photonics MLXX-EYFA-CW-SLM-P-TKS). The pump (probe) beam propagates along $z$ ($x$) axis and has the power of 15~W (0.3~mW) with $1/e^2$ waist diameter of about 20~mm (1~mm). The pump beam keeps the circularly polarized before entering  $^{3}\textrm{He}$ atomic cell, and the probe beam is circularly or linearly polarized to observe the orientation and alignment effect of the $F=3/2$ metastable states. The home-made pure $^{3}\textrm{He}$ (pressure: 0.6 Torr) cylindrical atomic cell (size: $\phi$50$\times$L70 mm$^3$ ) is located in the seven-layer magnetic shield, and is excited by a radio-frequency power source (50 MHz, 0.8~W) to continuously discharge and generate the metastable-state atoms. The solenoid generates the static magnetic field $B_0$ along $z$, and a set of helmholtz coil generates the oscillating magnetic field $B_{\rm M}$ along $y$. The digital processing system includes the PXI-4461 and PXI-4462 (resolution: 24-Bit, sampling rate: 204.8 kS/s) of the National Instruments, which is used for controlling the helmholtz coil and signal acquisition.

\begin{figure}
	\includegraphics[width=0.45\textwidth]{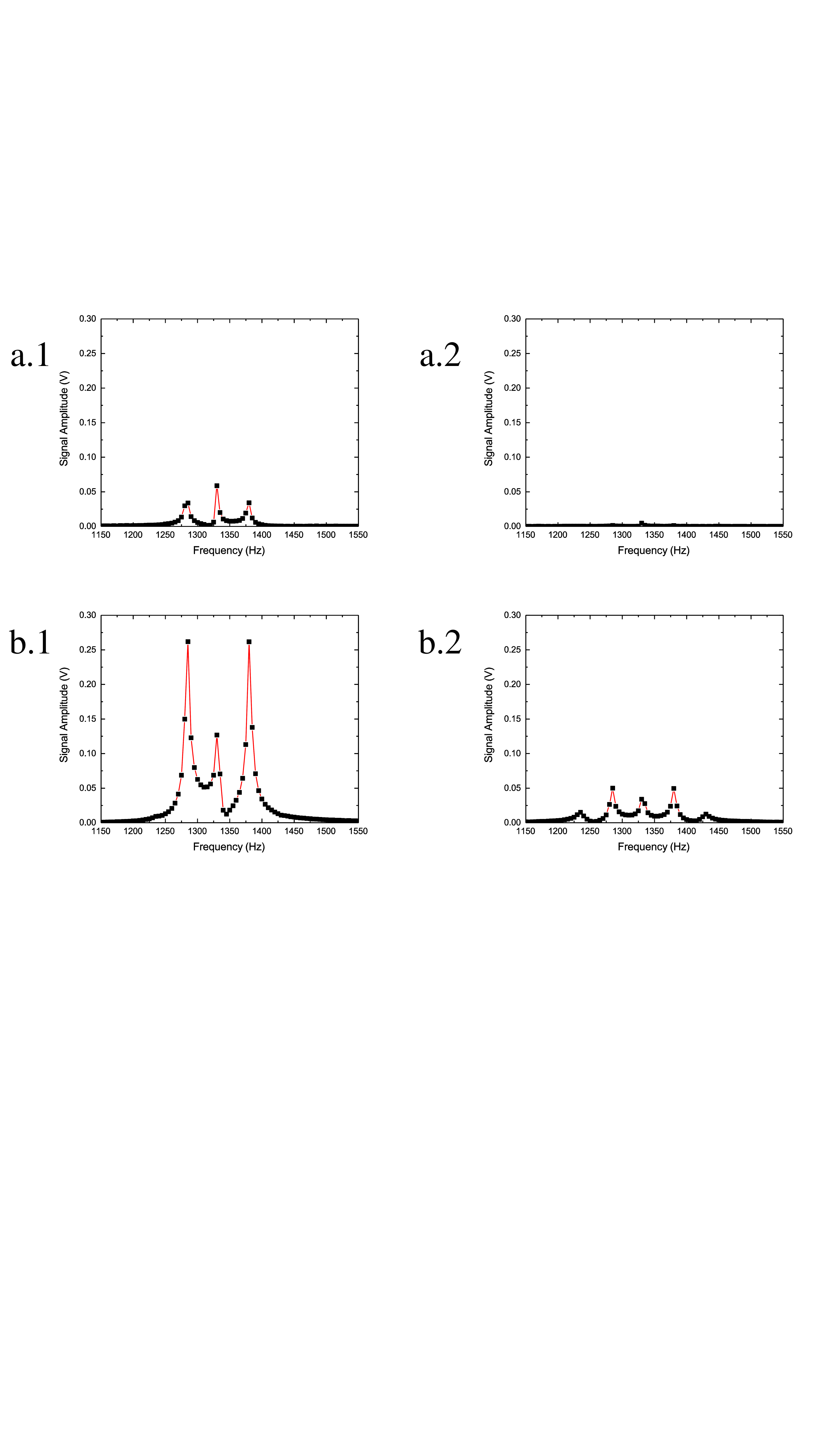}
	\caption{\label{fig:onoffpump} The frequency spectrum by detection with different optical polarization and different hyperfine structure states. Figures in first (second) row are detected by $C_8$ ($C_9$) line, and figures in left (right) column are detected by circularly (linearly) polarized light. The black squares are the experimental data, and the red lines are the fitting curves.
	}
\end{figure}

Figure~\ref{fig:onoffpump} manifests the MT and MQ spectra by detection of different optical polarization and different hyperfine structure states. The first row of the Fig.~\ref{fig:onoffpump} shows the detection with $C_8$ line, while the second row shows the $C_9$ ones. The pumping beam keeps the circularly polarized light, and the Fig.~\ref{fig:onoffpump}(a.1 and b.1) shows the circularly polarized probe while Fig.~\ref{fig:onoffpump}(a.2 and b.2) shows the linearly polarized probe. Comparing each figure in Fig.~\ref{fig:onoffpump}, the MT appears by the detection with circularly polarized light no matter which hyperfine structure state is probed, while the MQ appears by the linearly polarized probe and only in $F=3/2$ hyperfine structure state. Notice that the center frequency of the MT and MQ signal is the Larmor frequency of the ground state. We have reported a MT signal of ground state can be transferred through metastability exchange collisions (MECs) to the metastable states, which reveals that the signal of Fig.~\ref{fig:onoffpump}~(b.1) is the transferred MT of ground state. The different polarized probe beam indicates that the orientation and alignment effect of the metastable states atoms may be related to the MT and MQ signal, respectively.

\begin{figure*}
	\includegraphics[width=0.9\textwidth]{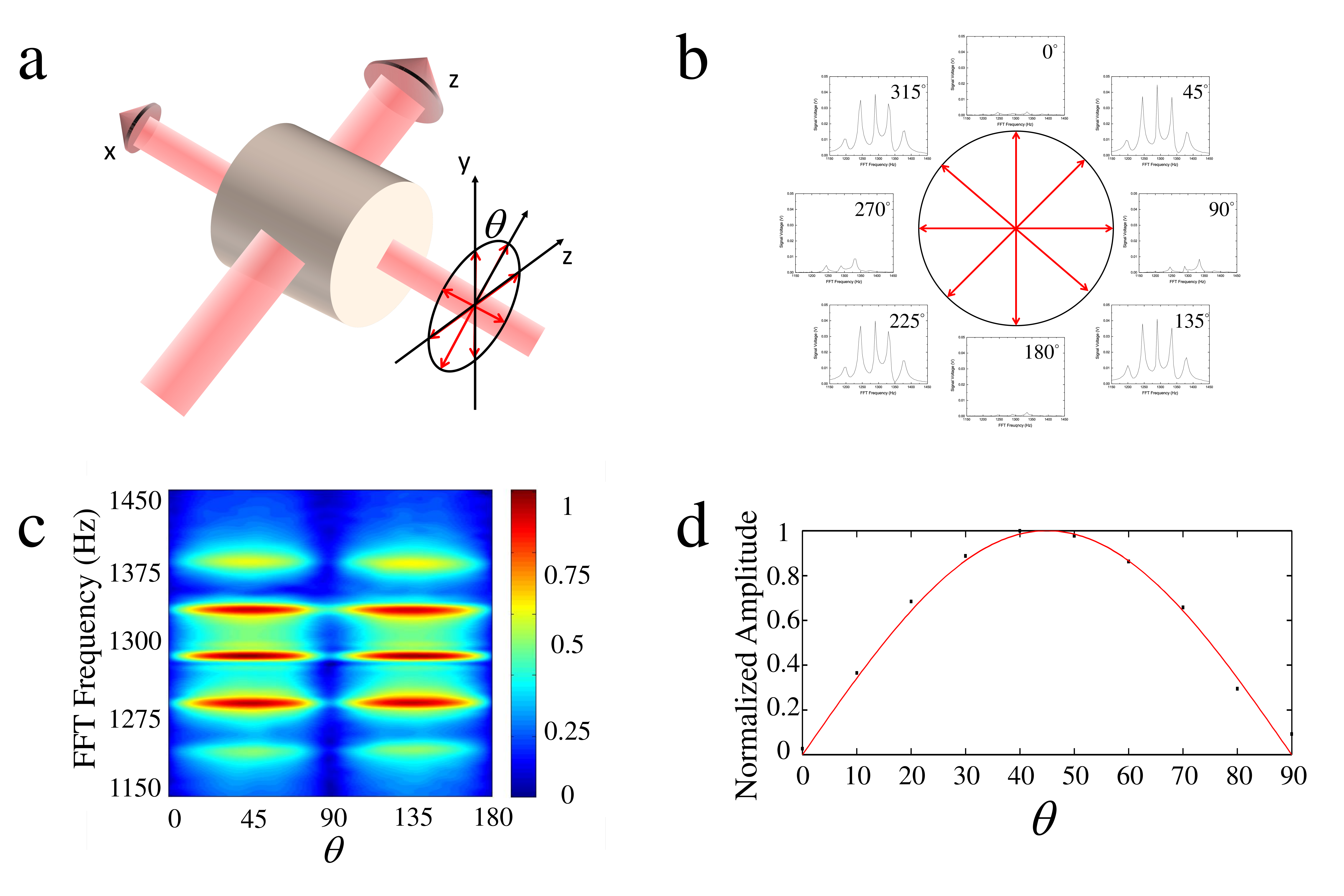}
	\caption{\label{fig:direction}The relationship between the MQ signal and direction of linearly polarized light. Figure(a) shows $\theta$ is the angle between the polarization of linearly polarized light and y axis. Figures (b) and (c) show the MQ signals in different $theta$. The black squares and red curve are experimental data and $\rm sin$$(2\theta)$ function in Fig.(d). The amplitude of static magnetic field is 5$\times 10^4$~nT.
	}
\end{figure*}

The circularly polarized pump beam (in z axis) and linearly polarized probe beam (in x axis) continuously interact with the atoms, and the frequency of oscillating magnetic field (in y axis) is set as $\omega =\omega_{\textrm{g}}$. The direction of the polarization is rotated in y-z plane shown in Fig.~\ref{fig:direction}(a), and the parameter $\theta$ is defined as the angle between the polarization and y axis. Changing $\theta$, the MQ signals will disappear at $\theta=0, 90$ (perpendicular or parallel the static magnetic field $B_0$) shown in Fig.~\ref{fig:direction}(b). The detailed relationship between MQ signal and direction of linearly polarized light is shown in Fig.~\ref{fig:direction}(c), which shows the amplitude of MQ each peaks is changed simultaneously with $\theta$, and the maximum of the amplitude appears at the condition $\theta = 45, 135$. Focusing on the amplitude of center peak, the relationship with $\theta$ is shown in Fig.~\ref{fig:direction}(d). The experimental data (black squares) are corresponding to the function $\rm sin$$(2\theta)$.

\begin{figure}
	\includegraphics[width=0.45\textwidth]{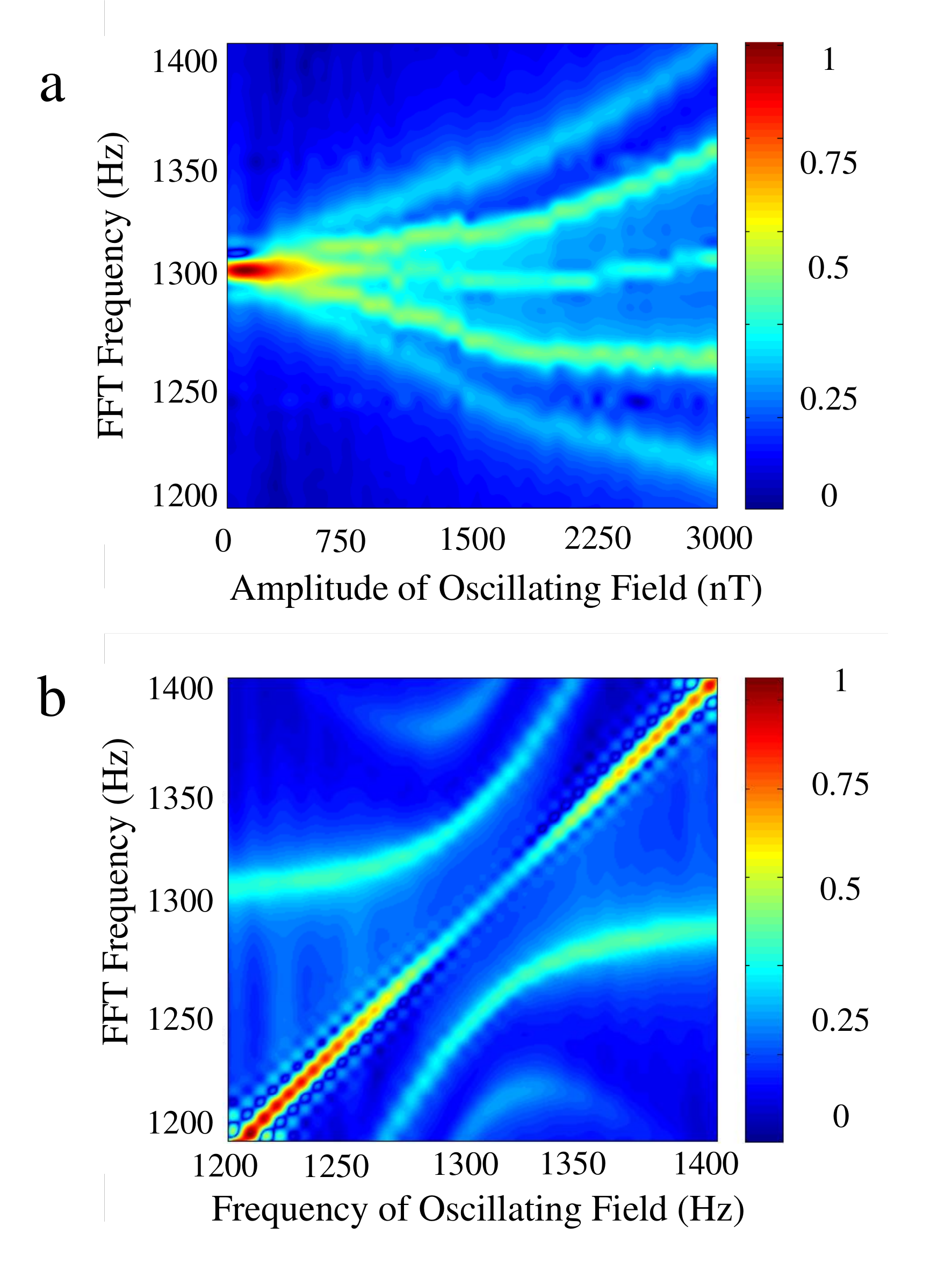}
	\caption{\label{fig:MTC8C9} Dependence of FFT spectrum on changing the amplitude (a) and frequency (b) of the oscillating magnetic field for experiment. The amplitude of oscillating magnetic field in is about 3000~nT, and the frequency of the oscillating magnetic field is at about 1.30~kHz.
	}
\end{figure}

With changing both the amplitude and frequency of oscillating magnetic field at the condition $\theta = 45$, the MQ signals are shown in Fig.~\ref{fig:MTC8C9}. The interval of each peak is the Rabi frequency $\Omega_R$, and the first order sidebands and second order sidebands appear together with changing the amplitude of oscillating magnetic field. $\Omega_R$ approximately behaves linearly with changing the amplitude of the resonant oscillating magnetic field $B_M$, which accords with $\Omega_R =(1/2)\gamma_{0}B_{\textrm{M}}$ and indicates a method of measuring the amplitude for the oscillating magnetic field. Figure~\ref{fig:MTC8C9}(b) shows the center peak and the first sidebands changing like MT signal with changing frequency $\omega_M$ of oscillating field, and the second order sidebands only appears at the $\omega_M \approx \omega_L$. The small sidebands near center peak appearing at the far detuning condition still need further researches. 

\begin{figure}
	\includegraphics[width=0.45\textwidth]{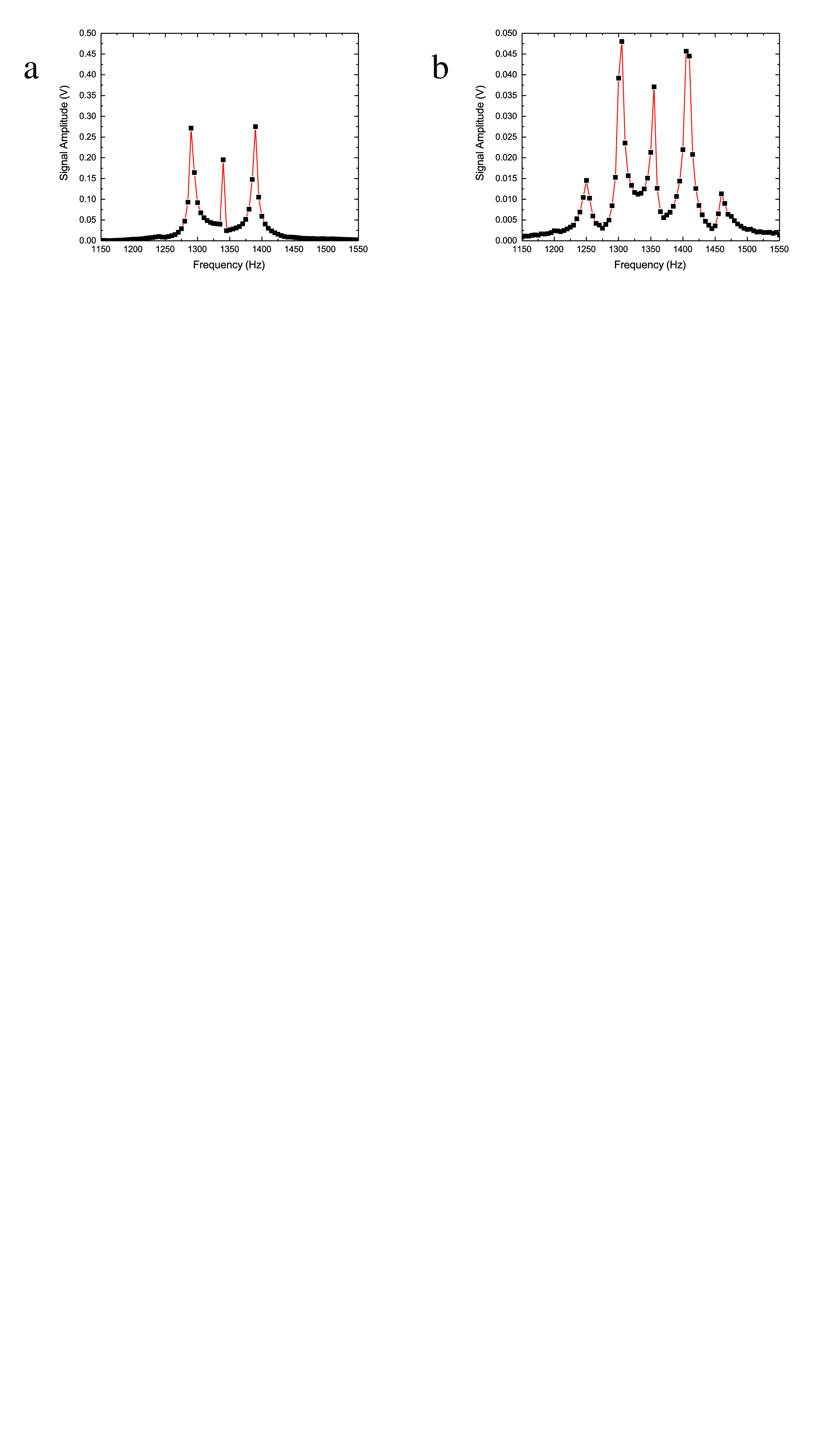}
	\caption{\label{fig:D0} The frequency spectrum of detection by different optical polarization with $D_0$ line of $^{4}\textrm{He}$. The detections by circular polarized light and linearly polarized light are shown in (a) and (b), respectively. The black squares are the experimental data, and the red lines are the fitting curves.
	}
\end{figure}

Depending on the experimental results, an possible explanation of the Mollow quintuplet manifests here. The physical picture of MT has been described as the dressed atom \cite{cohenbook}, and the MT transfer from the ground state to $F=1/2$ metastable state has been reported \cite{zhan2018observation}. The oscillating magnetic field drives the ground-state coherence of two Zeeman energy eigenstates or bare states $\vert$$\uparrow$$\rangle _{\textrm{g}}$ and $\vert$$\downarrow$$\rangle _{\textrm{g}}$. The new energy eigenstates or a series of dressed states are formed as the superposition states $\vert$$+$,$N$$\rangle_{\textrm{g}}$$=(1/\sqrt{2})$$($$\vert$$\downarrow$,$n$$\rangle_{\textrm{g}}$$+$$\vert$$\uparrow$,$n-1$$\rangle_{\textrm{g}}$$)$ and $\vert$$-$,$N$$\rangle_{\textrm{g}}$$=(1/\sqrt{2})$$($$\vert$$\downarrow$,$n$$\rangle_{\textrm{g}}$$-$$\vert$$\uparrow$,$n-1$$\rangle_{\textrm{g}}$$)$, where $n$ is the quantum number of oscillating field, $\vert$$\uparrow$,$n$$\rangle_{\textrm{g}}$ ($\vert$$\downarrow$,$n$$\rangle_{\textrm{g}}$) is the direct product state of atoms and the oscillating field and $N$ is the total number of the excitations in the system. Note that the energy interval $\hbar \Omega_R = (1/2)\hbar\gamma_{\textrm{g}}B_{\textrm{M}}$, where $B_{\textrm{M}}$ is the amplitude of the oscillating magnetic field. The MT can be transferred from the ground state to the metastable states has been reported, which reveals that the MECs between the dressed ground state and the undressed metastable states is a linear effect of the first order tensors of the density matrix $J_{+}$ and $J_{-}$. Furthermore, the MECs between the dressed ground state and the dressed metastable states create higher order nonlinear effect, and the MQ is the second order tensors of the density matrix like $J_{+}J_{z}$ and $J_{-}J_{z}$. The second order tensors $J_{+}J_{z}$ and $J_{-}J_{z}$ include two transitions, which are between $\Delta N = 1$ and $\Delta N = 0$ dressed states. The two transitions happen continuously, like that the Rabi oscillation modulates the MT, which change the frequencies of MT to the $\omega_{+}\pm \Omega_{\rm R}$, $\omega_{g}\pm \Omega_{\rm R}$ and $\omega_{-}\pm \Omega_{\rm R}$. The spectrum of the effect contain five frequencies $\omega_{g}+2\Omega_{\rm R}$, $\omega_{g}+\Omega_{\rm R}$, $\omega_{g}$, $\omega_{g}-\Omega_{\rm R}$ and $\omega_{g}-2\Omega_{\rm R}$ called Mollow quintuplet. In order to verify the explanation, the experimental results of $^{3}\textrm{He}$-$^{4}\textrm{He}$ hybrid cell is show in Fig.~\ref{fig:D0}. The metastable state of $^{4}\textrm{He}$ has three Zeeman states whose alignment effect also can be detected by linearly polarized light, and the spectrum gives a MQ signal.

We have observed the MQ signal in $F=3/2$ hyperfine structure state of $^{3}\textrm{He}$ atoms by detecting with linearly polarized light tuned to $C_9$ line. Comparing the MT and MQ signal at different polarization and center frequency of probe light, we give an possible explanation for MQ is related to the alignment of $^{3}\textrm{He}$ atoms, and the experimental results of $^{3}\textrm{He}$-$^{4}\textrm{He}$ hybrid cell give more evidence for the explanation. The difference between the MT and MQ signal reveals the demand of a new model for the MECs. The existing models of quintuplet spectrum are three levels system coupled two strong laser, two levels system coupled bimodality cavity, and two orthogonal polarization coupled one strong laser, but they cannot describe our experiments. We only give a phenomenological description to explain the generation of MQ, and the further theory needs more research. The frequency interval of the sidebands is linear with the amplitude of the resonant oscillating magnetic field, which satisfies $\hbar \Omega_R = (1/2)\hbar\gamma_{g}B_{\textrm{M}}$, and indicates the possible method of measuring the amplitude of the oscillating magnetic field. \\

\noindent \textbf{Funding.} This project is supported by National Natural Science Foundation of China (61571018, 61531003, 91436210); National key research and development program. \\

\noindent \textbf{Acknowledgment.} We thank for W.L., H.W. and H.d.W with experimental and technical assistance. 

\nocite{*}
\bibliographystyle{unsrt}

\end{document}